\documentclass[%
superscriptaddress,
 amsmath,amssymb,
 aps,
 prl,
 preprint,
]{revtex4-2}

\usepackage{graphicx}% Include figure files
\usepackage{dcolumn}% Align table columns on decimal point
\usepackage{bm}% bold math
\setcitestyle{super}
\usepackage{titlesec} % modify subsection
\titleformat{\subsection}[runin]
  {\normalfont\bfseries} % Formatting for the subsection title
  {}                     % Label
  {0pt}                  % Space between label and title
  {}                     % Before-code
\titleformat{\section}[hang]{\bfseries\large}{\thesection}{1em}{} 
\titlespacing*{\section}{0pt}{0.5em}{0.2em} % Adjust spacing: {left}{before}{after}

\usepackage{verbatim}
\usepackage{siunitx}
\usepackage[hidelinks]{hyperref}
\usepackage{comment}

\usepackage{xcolor}

\begin{document}

%affiliations 
\newcommand{\Lux}{Department of Physics and Materials Science, University of Luxembourg, 30 avenue des Hauts-Fourneaux, 4362 Esch-sur-Alzette, Luxembourg}
\newcommand{\FU}{Department of Physics, Freie Universit\"at Berlin,  Arnimallee 14, 14195 Berlin}
\newcommand{\LIST}{Materials Research and Technology Department, Luxembourg Institute of Science and Technology, 41 rue du Brill, L-4422 Belvaux, Luxembourg}
\newcommand{\IRG}{Inter-institutional Research Group Uni.lu–LIST on Ferroic Materials, 41 rue du Brill, L-4422 Belvaux, Luxembourg}
\newcommand{\BVO}{BiVO$_4$}
\newcommand{\Ag}{A$_\mathrm g$}
\newcommand{\Bg}{B$_\mathrm g$}

%\preprint{APS/123-QED}

%\title{Spectroscopic signatures of electron-polarons in monoclinic bismuth vanadate}
%\title{Resonant Raman signatures of excitonic polarons in monoclinic bismuth vanadate}
%\title{Resonant Raman signatures of excitonic polarons in a transition metal oxide: BiVO$_4$}
%\title{ Excitonic polaron mediated Raman scattering, the case of BiVO$_4$} %in a transition metal oxide: BiVO$_4$
%\title{Raman scattering enhanced by excitonic polarons in BiVO$_4$}
\title{Raman resonances mediated by excitonic polarons in BiVO$_4$ }
% Force line breaks with \\

\author{Georgy Gordeev}
\email{georgy.gordeev@uni.lu}
 \affiliation{\Lux{}}
 \affiliation{\IRG{}}
 
\author{Christina Hill}%
\affiliation{\Lux{}}
\affiliation{\IRG{}}
\affiliation{\LIST{}}
\author{Angelina Gudima}
\affiliation{\Lux{}}
\author{Stephanie Reich}
\affiliation{\FU{}}
\author{Mael Guennou}
\affiliation{\Lux{}}

\date{\today}% It is always \today, today,

\begin{abstract}
%excitons and localized polaronic state
%The excitons trapped by a polaron formation in \BVO{} can contribute to the resonant Raman cross-sections. To measure this effect, 
% Introduction sentence on excitonic polarons in BVO:
%Bismuth vanadate is a well-established host for polarons and excitonic polarons dominating the charge carrier transport in the material. 
% Excitonic polaron, exciton interacting with lattice vibrations  
%Excitons and polarons are quasi-particles that play a fundamental role in transition metal oxides, significantly influencing key physical properties such as light absorption and charge carrier transport. 
%bound together through Coulomb interaction
%In this work we exploit the intrinsic sensitivity of Raman spectroscopy to electron-phonon coupling in \BVO, a known host of excitonic and polaronic effects. 

%, a known host of excitonic and polaronic effects
Excitonic polarons are quasiparticles formed by a Coulomb-bound electron-hole pair with strong coupling to lattice vibrations. Despite high fundamental interest in excitonic polarons, the experimental investigation of these particles remains challenging. In this work, we exploit the resonant Raman effect to probe the excitonic polarons in bismuth vanadate. We track enhancement of Raman modes as a function of excitation energy and reveal two optical resonances: one inside the band gap at \SI{1.94}{\eV} and another one near the optical absorption edge at \SI{2.45}{\eV}. The high-energy resonance originates from free excitons, which exhibit a characteristic \SI{40}{\meV} anisotropy between polarizations parallel and perpendicular to the $c$ axis. Remarkably, the low-energy resonance shows no contrast in the optical absorption spectra. We attribute this resonance to an excitonic polaron formed through strong exciton–phonon coupling, making excitonic and excitonic-polaron Raman resonances similar in strength. We probe the energy level of the excitonic polaron and compare its coupling strength to the different vibrational modes. Our results establish resonant Raman spectroscopy as a unique and powerful tool for probing quasiparticles of polaronic and excitonic nature in oxide materials. 

%Raman spectra were recorded with 16 laser lines between 1.9 and \SI{2.6}{\eV} and analyze intensity variations of the Raman modes with \Ag{} symmetry at different energies.

%It manifests in the vibrations of vanadium and oxygen atoms where polaron localization occurs and the resonance energy matches theoretical predictions. The vibrational modes couple to the excitonic polaron with different efficiency determined from resonant Raman profiles. 

\end{abstract}

\maketitle
%\newpage
%\tableofcontents
%\section{Introduction}

%\subsection{Excitons, polarons and co}
%The interplay between lattice polarization and Coulomb forces can lead to the formation of exceptional states, redefining the physical properties of materials.
%, also known as polaronic excitons,

% While bare polarons can be detected by single-particle techniques, the EPs are more elusive.
% A defining characteristic of these EPs is often considered a
%in which both electrons and holes attract phonon clouds across the entire Brillouin zone. While being an important signature, the EP emission spectra alone do not

A bound electron-hole pair forms an exciton, a neutral quasiparticle that modifies the absorption band-edge. When either an electron or a hole attracts a local phonon cloud, a polaron is formed, altering transport phenomena.\cite{Austin1969,Bosman1970} Remarkably, a simultaneous strong interaction between an exciton and a phonon results in an excitonic polaron (EP).\cite{Toyozawa1968, Iadonisi1983, Dai2024, Dai2024b} The EP intertwines optical and electrical transport properties, relevant for photo-catalysis and photo-voltaic applications.\cite{Dai2024, Wiktor2018}
Despite tremendous interest, the unambiguous detection of EPs remains challenging. Most works focus on the emission properties, where strong exciton-phonon coupling produces a new emission line labeled as a self-trapped exciton (STE).\cite{Dai2024b} The emission of EPs is dominated by the post-absorption relaxation processes and is often represented by a broad emission peak associated with a significant Stokes shift.\cite{li_self-trapped_2019,Li2020, abfalterer_colloidal_2020, bai_temperature-dependent_2024, moslinger_competing_2025} %However a broad, emission spectrum can sometimes originate from deep, shallow defects and therefore, mimic EP Stokes-shifted emission line. 
The Stokes-shifted emission however, can be confused with other phenomena and does not directly yield EP key characteristics, such as formation energy,\cite{Pham2020, Kweon2015, Wiktor2018}, formation lifetime, and phonon modes that participate in EP formation. Novel experimental methods sensitive to exciton-phonon coupling are necessary to capture the EP formation and resolve the phonon branches dressing EPs. \par
Monoclinic bismuth vanadate (\BVO{}) is a well-established host for excitons\cite{Das2017} and polarons.\cite{Pham2020, Kweon2015, Kweon2013, Liu2019, Liu2015, Wiktor2018} It has been shown that excess holes preferentially form polarons that are delocalized across multiple lattice sites\cite{Kweon2013, Wiktor2018, liu_hole_2020}, classifying them as large polarons. These large hole polarons exhibit moderate drift mobility under electric field. In contrast, excess electrons are likely to form polarons that are highly localized within a single unit cell, referred to as small polarons.\cite{Kweon2015, Wiktor2018, aiga_electronphonon_2013, rettie_unravelling_2016} Due to their strong localization, these small electron polarons have extremely low mobility and significantly reduce the charge carrier conductivity of \BVO{}. The optical absorption edge\cite{Das2017} in \BVO{} is dominantly defined by excitons, and these excitonic effects remain substantial up to room temperatures.\cite{wiktor_comprehensive_2017} The coexistence of strong Coulomb interactions and significant coupling to lattice vibrations suggests the formation of excitonic polarons, verified both experimentally\cite{Zhang2022} and computationally\cite{moslinger_competing_2025}. With an optical absorption edge in the visible light range at approximately \SI{2.50}{\eV}\cite{cooper_indirect_2015}, the various excitonic and polaronic states become relevant for solar-light photo-induced effects. This makes \BVO{} an ideal candidate for identifying and characterizing excitonic and polaronic states via resonant Raman spectroscopy.\par

% The various exitonic and polaronic states in \BVO, have been evidenced using different techniques such as photoluminescence\cite{fernandez_strain-induced_2023}, transient absorption, and time-resolved THz spectroscopy \cite{Zhang2022}.
%This method also has the potential to reveal the characteristic features of polaronic and excitonic in-gap states.\par

%The FE can efficiently interact with photons, given its high transition dipole moment, but has a moderate coupling to phonons. On the other hand, the EPs have comparably smaller dipole moments, since the overlap of localized and delocalized wavefunctions is faint. For that reason, the EP states rarely appear in linear optics, yet they can be rather dominant when phonon scattering is involved within a Raman scattering process. 

In this work, we implement resonant Raman scattering to investigate excitonic polarons in \BVO{}. The \Ag{} modes are excited wtih 16 laser energies, across and below the band gap. We measure the Raman cross-section as a function of excitation energy. Resonances occur when the excitation approaches either the optical band gap or a new in-gap transition. We discuss the origin of the in-gap resonance and attribute it to the excitonic polaron. Notably, this EP transition is invisible in linear optical absorption but reappears at a different energy in form of emission from a self-trapped exciton. Finally, we discuss the nature and the properties of the observed EP and establish its new distinctive signature, representing a strong Raman resonance accompanied by negligible optical absorption.\par

 \begin{figure*}
  \centering
  \includegraphics[width=\linewidth]{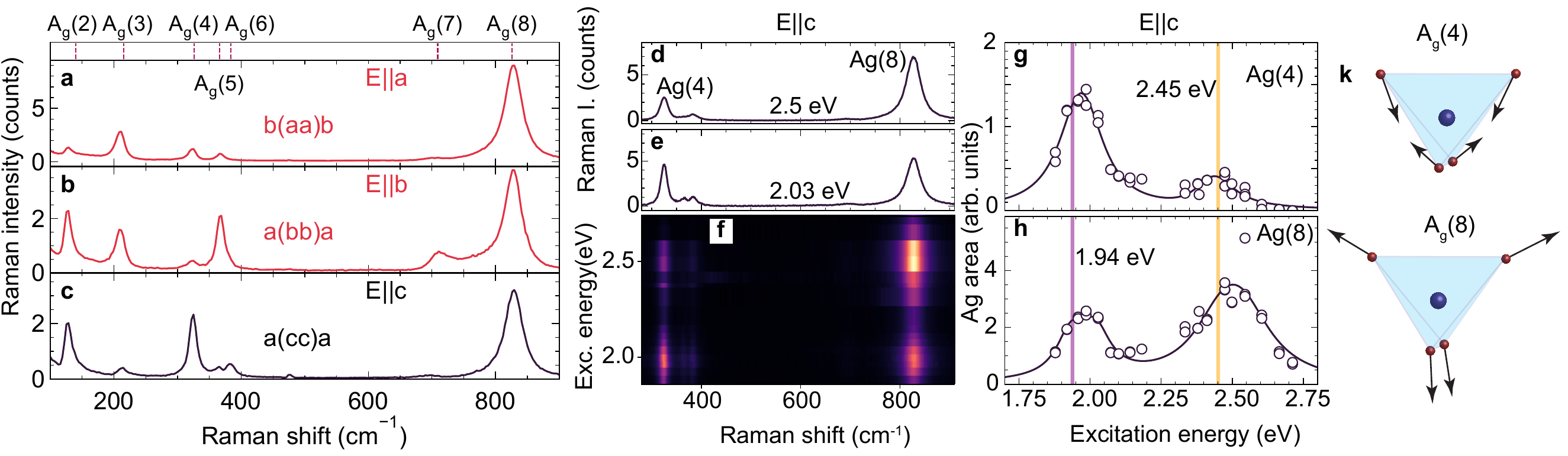}
  \caption{Resonant Raman spectroscopy applied to the \BVO{} single crystal. Raman spectra acquired with \SI{2.3}{\eV} excitation for incident and scattered photons polarized along \textbf{a} $E||a$, \textbf{b} $E||b$, and \textbf{c} $E||c$.
  \Ag{} modes with $E||c$ \textbf{d} excited near the band gap energy at \SI{2.5}{\eV} and \textbf{e} inside the gap  at \SI{2.03}{\eV}. \textbf{f} Resonant Raman map measured with $E||c$. Resonant Raman profile of \textbf{g} the \Ag(4) and \textbf{h} the \Ag(8) mode, where symbols represent experimental data and solid lines are fits using Eqs. S1 and S2. Vertical lines indicate the optical transition energies extracted from the fit. \textbf{k} Eigenvectors of the \Ag(8) mode and \Ag(4) mode within a VO$_4$ tetrahedron.
 % , selecting diagonal tensor components $R_{\alpha \alpha}$ of the Ag modes. 
  }
  \label{FIG:Ram}
\end{figure*}

%The EP state is most likely formedd by a small Holstein electron polaron in \BVO{}. The crystal structure of \BVO{} is shown in Figure \ref{FIG:scheme}a. Vanadium atoms are surrounded by oxygen tetrahedra. At room temperature \BVO{} has a C2/c symmetry with lattice parameters (in the non-standard but more convenient I2/a setting) $a = \SI{5.186}{\angstrom}$ , $b = \SI{5.084}{\angstrom}$, and $c = \SI{11.69}{\angstrom}$, $\gamma = \SI{89.61}{\degree}$.\cite{Roth1963, Choh2016} The small Holstein polarons are predicted to occur at the VO$_4$ tetrahedra. The excitonic transition originate from $3d$ electrons in vanadium coupled to holes in oxygen from $2p$ orbitals.\cite{Das2017,Zhao2011} The polaronic wave function as calculated in~\citet{Pham2020} is depicted in Figure \ref{FIG:scheme}b, it is formed by the $d_{z^2}$ electrons of vanadium and anti-symmetric stretching of the VO$_4$ bond, as schematically shown in Figure \ref{FIG:scheme}b. The polaronic state is localized and typically has lower energy compared to the delocalized electron by $2W_\mathrm h$ where $W_\mathrm h$ is the polaron hopping energy, see Figure \ref{FIG:scheme}c. Both electron and polaron can couple to the valence band holes yielding a formation of excitonic and EP states, respectively, shown by dashed lines in Figure \ref{FIG:scheme}c. The energy difference between a hole-coupled and hole-uncoupled states is called exciton (EP) binding energy $E_B$. \par

\section{Results and discussion}
%With \SI{2.3}{\eV} excitation, the highest intensity is provided by t
% corresponding to the breathing of the VO$_4$ tetrahedron; see Figure \ref{FIG:Ram}d
At room temperature monoclinic \BVO{} has a $C2/c$ symmetry with lattice parameters (in the non-standard $I2/a$ setting) $a = \SI{5.186}{\angstrom}$, $b = \SI{5.084}{\angstrom}$, and $c = \SI{11.69}{\angstrom}$, $\gamma = \SI{89.61}{\degree}$.\cite{Roth1963, Choh2016} The Raman-active  vibrations of \BVO{} are well characterized experimentally \cite{avakyants_inelastic_2004, Porres2018} and have been identified through DFT calculations.\cite{Porres2018} Group-theory analysis predicts 18 Raman-active modes, consisting of 8 modes of \Ag{} symmetry and 10 modes of \Bg{} symmetry. In this work, we focus on the \Ag{} modes; they are activated when the incident and scattered polarization are aligned. The \Ag{} modes originate from the motion of oxygen and vanadium atoms, where small polarons form.\cite{Pham2020} Figure \ref{FIG:Ram}a–c shows the Raman spectra of \BVO{} for light polarized along three different crystallographic directions. Among all \Ag{} modes, the \Ag(8) mode at approximately \SI{826}{\per\cm} exhibits the strongest intensity in all three polarization configurations. In the following, we focus on \Ag(8) and \Ag(4) modes because of its dominant intensity and strong polarization-dependent Raman response. For instance, the \Ag(4) mode at \SI{326}{\per\cm} becomes comparably strong for the electric field of the laser light $E$ oscillating $\parallel c$, corresponding to $a(cc)a$ in Porto’s notation, as shown in Figure \ref{FIG:Ram}c. \par

\begin{comment}
 \begin{figure}
    \centering
    \includegraphics[width=0.49\textwidth]{./Fig/BVO_RR_Fig2.pdf}
    \caption{Resonant Raman spectroscopy of the \Ag modes with a(cc)a polarization. a) Raman spectrum excited with \SI{2.45}{\eV}, near the band gap energy, and b) Raman spectrum excited below the band gap at \SI{1.9}{\eV}. c) Resonant Raman map d,e) Resonant Raman profile of the \Ag(4) and \Ag(8) modes, symbols represent experimental data and lines are fits by Eq. S1. Vertical lines represent the transition energies of the polaron and electron states, shifted from the transitions by $E_{ph}/2$.} %Ag(8) = Ag(3) schellite Ag(4) = Ag(2)+Bg(3)
    \label{FIG:RR_hor}
\end{figure}   
\end{comment}
%fig3

\subsection{Raman resonances of the optical many-body states.} At first, we study in detail the  optical resonances for case when $E\parallel c$. Two Raman spectra are exemplified in Figure~\ref{FIG:Ram}d and \ref{FIG:Ram}e, excited with 2.50 and \SI{2.03}{\eV}, respectively. In this polarization configuration, the \Ag(8) and \Ag(4) modes exhibit the strongest intensities. The intensity ratio of these two modes varies with the excitation energy. At \SI{2.03}{\eV}, their amplitudes are almost equal, whereas at \SI{2.50}{\eV} the \Ag(8) mode dominates. The systematic variation in Raman intensity as a function of excitation energy is most evident in a two-dimensional resonant Raman map shown in Figure~\ref{FIG:Ram}f. In this map, the vertical axis represents the excitation energy, whereas the horizontal axis corresponds to the Raman shift, and the color scales with Raman intensity. Data were collected using 16 laser energies ranging from 1.87 to \SI{2.71}{\eV}. The \Ag(4) and the \Ag(8) modes are primarily visible. Along the excitation-energy axis, two maxima appear in the high- and low-energy ranges, signaling the presence of two optical resonances at approximately \SI{2.0}{\eV} and \SI{2.5}{\eV}.\par

The two observed Raman resonances yield a similar intensity despite different underlying nature of optical transitions. We quantify their parameters using resonant Raman profiles shown in Figure \ref{FIG:Ram}g. In such a profile, the y-axis represents the Raman cross section of the investigated Raman mode, while the x-axis corresponds to the excitation energy. Figure \ref{FIG:Ram}g shows the resonant Raman profile of the \Ag(4) mode, displaying two resonances at which \Ag(4) peak area reaches its maximum. The data fits by Eqs. S1 and S2 and represented by the solid line in Figure \ref{FIG:Ram}g. The model yields two optical transitions energies at \SI{2.40}{\eV} and \SI{1.94}{\eV}. We systematically observe two resonances near these energies in all profiles of the \Ag{} shown in Figure S2. The high-energy resonance occurs at the absorption edge\cite{Hill2020} and can be attributed to free delocalized exciton (FE). However the low-energy resonance at \SI{1.94}{\eV} is non-trivial and occurs well inside the gap. It may be preliminarily attributed to the EP state. To confirm this interpreation we next compare the optical response of two resonances and their coupling strength to different phonon modes.\par

%Local symmetry within the VO$_4$ tetrahedron plays an important role in EP-phonon coupling. For example, the \Ag(8) mode couples less efficiently with the EP state, compared to FE, whereas the \Ag(4) shows an opposite behavior, see Figure \ref{FIG:Ram}g,h. 

The small polaron wavefunction is confined within the VO$_4$ tetrahedron,\cite{Pham2020} such that the local symmetry of the vibrational modes governs the coupling strength. While the \Ag(4) mode responds strongly to the low-energy optical transition, the resonant Raman profile of the \Ag(8) mode shows a stronger resonance for the high-energy optical transition, see Figure \ref{FIG:Ram}h. These modes correspond to the motion of oxygen atoms within the different local symmetry. Locally, the \Ag(8) mode appears to be fully symmetric, see Figure \ref{FIG:Ram}k. All four oxygen atoms perform breathing-like vibrations by symmetrically changing only the V-O bonds length, inducing non-polar vibration. In contrast, the \Ag(4) mode has more of a bending-like character, with two top and bottom oxygen atoms moving upwards and downwards simultaneously, where the angles between the V-O bonds change, see eigenvectors in Figure \ref{FIG:Ram}k. We believe that a distinct vibrational character of the \Ag(4) mode causes a different resonant Raman response compared to the \Ag(8) mode.  Thus, resonant Raman profiles efficiently compare coupling to phonons of the excitonic and EP states. We can further analyze EP-photon coupling strength using linear optical spectroscopy.\par

 \begin{figure}
  \includegraphics[width=0.99\linewidth]{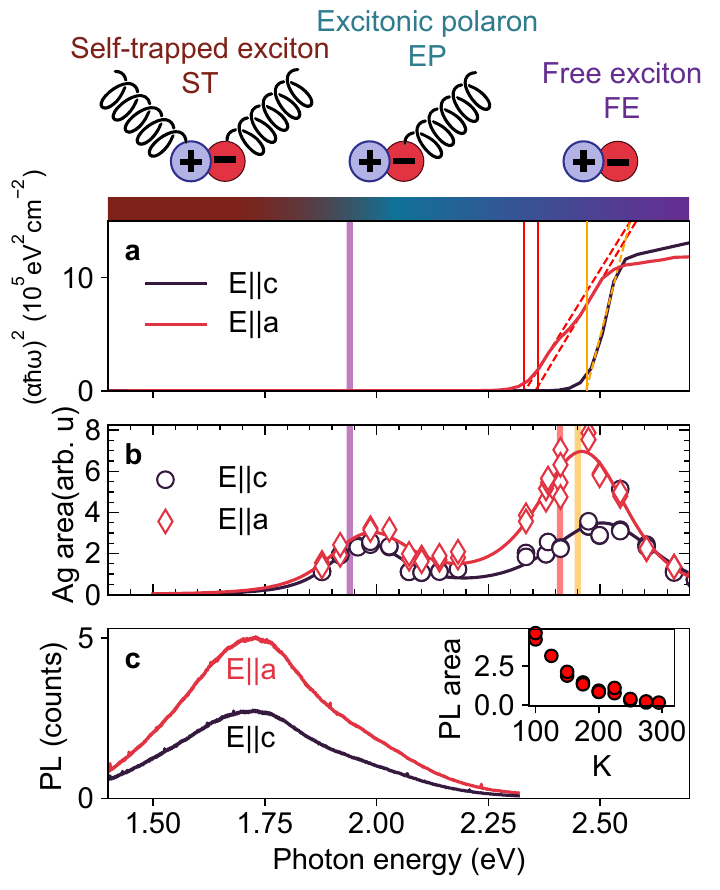}%0.49
  %BVO_RR-fig_polarized-v2_Fig24prl
  \caption{Optical signatures of excitonic and EP states. \textbf{a} Tauc plot showing anisotropy of the optical absorption edges for light polarized along the crystallographic $c$ and $a$ axes. \textbf{b} Resonant Raman profiles of the \Ag(8). The yellow (red) vertical line indicates the transition energy of free excitons for $E||c$ ($E||a$) in the energy range of the optical band gap. The vertical purple line marks the energy of the EP state, which is only observed in resonant Raman spectroscopy. \textbf{c} Photoluminescence from self-trapped excitons with the maximum at \SI{1.74}{\eV} excited with a \SI{2.8}{\eV} laser. The inset shows the evolution of the PL intensity as a function of temperature.}
  \label{FIG:OPT}
\end{figure}

%Anisotropy in the lowest optical band gap energies visible both in transmission and reflectivity spectra, shown in Figure \ref{FIG:OPT}a,b.
%The interaction between charge carriers and the lattice gives rise to a variety of many-body states, each associated with distinct optical transitions as depicted in Figure \ref{FIG:scheme}b. The free exciton (FE) represents the highest-energy transition and consists of a delocalized electron-hole pair bound by Coulomb attraction. In contrast, the excitonic polaron features a delocalized hole and an electron dressed by lattice distortions, forming a polaronic state stabilized inside the electronic band gap, see Figure \ref{FIG:scheme}a. The lowest-energy configuration is the self-trapped exciton (STE), in which both electron and hole are localized by strong coupling to the phonon field, leading to full lattice relaxation. While FE and excitonic polarons participate in absorption processes and can mediate resonant Raman scattering, the ST exciton is optically active only in emission, with no corresponding direct absorption pathway.\par
%fig3
\begin{figure}
    \centering
    \includegraphics[width=8cm]{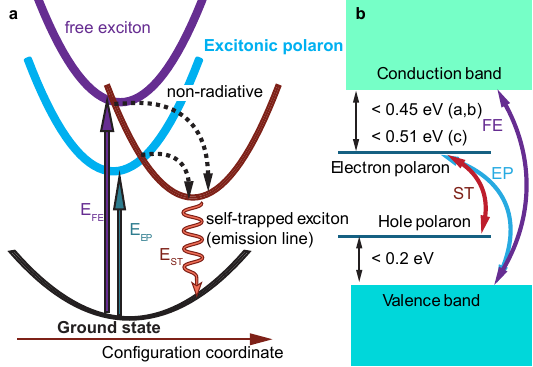}
    %BVO_RR-fig_polarized-v2_Fig24prl
    \caption{Energy level diagram and formation energies of EP. \textbf{a} Many-body picture with free excitons (purple), excitonic polarons (blue), and self-trapped excitonic states (red) in BiVO$_4$. Direct up-pointing arrows indicate optical absorption, tilted arrows indicate non-radiative scattering, and the wavy arrow represents the emission from the self-trapped excitonic state into the ground state. \textbf{b} Energy diagram with the in-gap energy levels of electron and hole polarons in \BVO{}. 
    }
    \label{FIG:scheme}
\end{figure}
%, formed by localized electrons and delocalized holes
%The EP and FE states can be discriminated by their inherently different optical response. 

\subsection{Optical response of free excitons and excitonic polarons.} We carry out linear optical spectroscopy to investigate the high- and low-energy Raman resonances. We quantify the optical gaps using the Tauc plot shown in Figure \ref{FIG:OPT}a, where $(\alpha \hbar \omega)^n$ is plotted versus $\hbar \omega$. We choose $n=2$ which is most appropriate for direct allowed interband transitions in crystalline materials.\cite{Klein2023} The absorption coefficient $\alpha$ is calculated as $\frac{1}{d}log(\frac{1-R}{T})$. The linear fit to $\alpha=0$ yields an optical gap $E^c_{opt} = \SI{2.47}{\eV}$ for $E||c$  and $E^{a,b}_{opt} =  2.35\pm 0.02$ for $E\perp c$, with an optical anisotropy of approximately $ \SI{120}{m\eV}$, similar to previous reports that rely on linear optics.\cite{Hill2020, cooper_indirect_2015, Das2017} However Tauc plots are limited in terms of absolute numbers and overestimate the anisotropy of the band gap.\cite{Klein2023} The gap anisotropy is measured more accurately by resonant Raman spectroscopy and yields \SI{40}{\meV}, see Figure \ref{FIG:OPT}b. The energy of the optical gaps matches the high-energy resonances at $\SI{2.45}{\eV}$ for $E||c$  and at $\SI{2.41}{\eV}$ for $E\perp c$, as shown in Figure  \ref{FIG:OPT}b . The anisotropic optical band gap near \SI{2.41}{\eV} is confirmed to be the free exciton, remaining bound at room temperature.\cite{Das2017,Zhang2022} The FE manifest regularly in optical and resonant Raman spectroscopies: it shapes the absorption band-edge and producing the Raman resonance at the same energy. \par
%From this point onward, we refer to the free exciton energy $E_{FE}$ rather than the absorption edge.\par

We now test the nature of the EP resonance at 
\SI{1.94}{\eV}, finding several observations consistent with our interpretation. Compared to the FE it is practically undetectable in linear optical spectroscopy, see purple line in Figure \ref{FIG:OPT}a. We as well find no signature in the reflectivity spectra shown in Figure S3. Indeed the interaction of EP with light is expected to be weak since the EP state is highly localized and energetically narrow.\cite{Dai2024b} Another important aspect is the exceptionally strong EP–phonon coupling observed in the resonant Raman profiles that have similar intensity with FE Raman resonances. The Raman intensity scales as $M_{EP-photon}^4$ and $M_{EP-phonon}^2$, where $M$ is the matrix element in SEq. (2). It requires quadratically stronger EP-phonon coupling to compensate the for weak EP interaction with photons and produce Raman resonances of comparable strength. Indeed, a strong EP-phonon coupling is expected for the EP formation, where charges are dressed by a phonon cloud. \par

%Further, the EP state well satisfies the energy criteria of many-body states incorporating polaronic and excitonic effects. \par

The observed EP state is most likely formed by a localized electron polaron and a delocalized hole from the valence band. In \BVO, electron polarons are expected to lie between \SI{0.3}{\eV}\cite{Kweon2015, rettie_unravelling_2016}, \SI{0.5}{\eV}\cite{Pham2020} and \SI{0.88}{\eV}\cite{Wiktor2018} below the conduction band, which coincides with the energy region where we observe the lower-energy Raman resonance. A hole-based EP, where localized hole couples to a delocalized electron, can be excluded because hole polarons are predicted to lie only $\approx$\SI{100}{meV}\cite{Kweon2013, ziwritsch_direct_2016,Wiktor2018, liu_hole_2020} above the valence band, far too deep inside the band gap compared to the experimental resonance energy. In addition, the expansion of hole polarons in monoclinic \BVO{} is large compared to the unit cell. The EP-phonon coupling for large polarons is predicted to be small at the $\Gamma$ point, where the Raman process occurs.\cite{Dai2024b} Another possible origin could be a defect-induced state, but in \BVO{}, point defects such as oxygen vacancies have been shown to lead to the formation of polarons.\cite{Osterbacka2022, kim_simultaneous_2015, seo_role_2018}

While hole-polarons weakly contribute to Raman scattering they can still form self-trapped excitons via finite momentum phonons, defining new photoluminescence line (PL). We therefore take a closer look at the emission properties of the \BVO{} crystal. In Figure \ref{FIG:OPT}c we show the PL emission peaking at \SI{1.74}{\eV}. Previously, this emission has been attributed to the defect-bound excitons, yet it's intensity is almost insensitive to hydrogen passivation.\cite{Cooper2016} Instead, we interpret this as a classical self-trapped exciton: its emission has a Stokes shift compared to EP and FE states. The emission intensity increases at low temperature, see the inset in Figure \ref{FIG:OPT}c, consistent with self-trapped exciton interpretation and previous reports in thin films.\cite{Cooper2016} The lattice trapping is more stable at low temperatures and higher STE emission is expected. Further, the full width at half maximum of this peak is twice as broad \SI{0.5}{\eV} as for the EP and FE states with 0.2-\SI{0.3}{\eV}, caused by the finite-momentum phonons, another key property of the self-trapped excitons. In \BVO{} it is clearly apparent that the EP and self-trapped excitons represent two distinct states, despite both being formed by a strong exciton-phonon coupling.\par

%When the Holstein polaron is hopping along the crystal axis it must overcome the energy of the local potential well $W_{h}$.\cite{Natanzon2020} For $W_h$ estimation we approximate the binding energies of exciton and EP to be the same. 

\subsection{Formation energies of many-body states in \BVO{}.} Figure \ref{FIG:scheme}a visualizes the optical absorption into free excitons and excitonic polarons while the emission occurs from the self-trapped excitonic states. From our resonant Raman and the PL measurements we can estimate the upper bound of hole- and electron-polarons formation energies $\Delta H_f$, assuming that the electron-hole binding energy of the localized EP states exceeds the binding energy of delocalized FE.\cite{Kshirsagar2024} In the main crystallographic directions, we find the same EP energy $E_{EP}$ \SI{1.94}{\eV}, see purple line in Figure \ref{FIG:OPT}a,b. The electron-polaron formation energy is estimated as $\Delta H_{f,el-pol} < (E_{FE}-E_{EP})$. For the $c$ axis we find $\Delta H_{f,el-pol}^c < \SI{0.51}{\eV}$, whereas for the $a$,$b$ axes we find $\Delta H_{f,el-pol}^{a,b}< \SI{0.45}{\eV}$, see Figure \ref{FIG:scheme}b. These values approach well the first-principles calculations from \citet{Kweon2015} and \citet{Pham2020}. Anisotropic electron-polaron hopping is driven by the variation of the formation energy and is more probable towards $a,c$ directions, as previously reported by facet-selective charge accumulation.\cite{Li2013, Xi2010, Wang2011} Similarly, we can estimate the upper limit for the hole-polaron formation energy $\Delta H_{f,hole-pol}$ by $(E_{EP}-E_{ST})$. The energy of the hole-polaron can be found at \SI{0.2}{\eV} above the valence band, see Figure \ref{FIG:scheme}b. \par

In summary, we implemented resonant Raman spectroscopy as a unique tool to distinguish and measure excitonic polarons. The polarization-resolved Raman profiles of the \Ag{} modes in \BVO{} reveal two well-separated resonances: a high-energy feature at 2.41–\SI{2.45}{\eV}, associated with free excitons and tracking the anisotropic optical absorption edge (\SI{40}{\meV}), and a sub-gap resonance at \SI{1.94}{\eV}, which we assign to excitonic polarons enabled by strong EP–phonon coupling. These findings demonstrate that resonant Raman spectroscopy provides direct and selective fingerprints of excitonic polaron formation, offering an experimental bridge to emerging theoretical descriptions.\cite{Dai2024, Dai2024b} Beyond this system, EP-mediated Raman processes enable access to a broad class of polaronic quasiparticles, including defect-bound \cite{Setvin2014}, spin \cite{Huang2017}, and surface polarons \cite{Kang2018}. Resonant Raman spectroscopy thus emerges as a powerful and versatile probe of polaronic excitations in semiconducting oxides.

%The hallmarks of excitonic polarons provide a new reference point from modern theories.
%In \BVO{} a large hole-type polaron can be formed around a bismuth atom in \BVO{}. The corresponding signatures would be contained in the \Bg{} modes that mostly involve Bi-O vibrations at the energies closer to the optical band gap. 
\section{Data availability}
The data were deposited online and will be published after the embargo period. DOI
\href{10.5281/zenodo.18679516}{10.5281/zenodo.18679516}
\section{Methods}
\subsection{Sample.} For this purpose, we used a single crystal purchased from SurfaceNet GmbH and grown by the Czochralski method. 
\subsection{Raman spectroscopy.} The Raman measurements were done on a Horiba T64000 microspectrometer with Rayleigh light rejected by long-pass filters and Raman signal dispersed by 900 grooves per mm grating. The laser light was produced by a Coherent 70C Ar-Kr and Radiant Dye dye lasers. The spectral intensities are normalized to the \Ag{} mode of a  CaF$_2$ crystal, which exhibits a constant Raman cross section in the visible wavelength range.
\subsection{UV-Vis spectroscopy.} Reflectivity and transmission were measured in \SI{5}{\nm} steps with light polarized parallel and perpendicular to the $c$ axis using a Lambda 1050+ uv/vis/nir Perkin-Elmer spectrometer; see Figure S3.\par 
\subsection{Photoluminescence.} The sample was excited with a He-Cd \SI{441.3}{\nm} laser line and re-emitted light collected and processed with a Renishaw spectrometer, equipped with 300 grooves per nm grating and Silicon CCD. For cryogenic temperatures, an Oxford Instruments Microstat was used.\par
%Discuss the changes of the small and large polarons after phase transitions

%exctionic polarons
%Hole polarons, defect bound polarons, Hybrid polarons, Bg modes with Bismuth involved

%\nocite{*}

\section{Acknowledgments}
\begin{acknowledgements}
 GG, MG, and CH acknowledge funding from the Fond National de la Recherche under Project PRIDE/15/10935404. This research was  funded  in  whole, or  in  part,  by  the  Luxembourg  National  Research Fund (FNR), grant reference [INTER/MERA20/14995416/SWIPE/AWORD]. For the purpose of open access, and in fulfillment of the obligations arising from the grant agreement, the author has applied a Creative  Commons Attribution  4.0  International  (CC  BY  4.0) license  to  any  Author Accepted Manuscript version arising from this submission. The authors are grateful to J. Iniguez, H-J Zhao, D. Vincent and X. Rocquefelte for preliminary calculations of \BVO{} phonon modes and eigenvectors. \\   
\end{acknowledgements}

%C.H. acknowledges funding from the Fond National de la Recherche (FNR) under Project No. PRIDE/15/10935404 \\
%Christina -> PRIDE MASSENA. Georgy -> M-ERA.NET SWIPE. The authors are grateful to J. Iniguez, H-J Zhao, D. Vincent and X. Rocquefelte for preliminar calculations of BVO phonon modes and eigenvectors. 

\bibliography{main}% Produces the bibliography via BibTeX.

\end{document}

%affiliations 
\newcommand{\Lux}{Department of Physics and Materials Science, University of Luxembourg, 30 avenue des Hauts-Fourneaux, 4362 Esch-sur-Alzette, Luxembourg}
\newcommand{\FU}{Freie Universit\"at Berlin, Department of Physics, Arnimallee 14, 14195 Berlin}
\newcommand{\LIST}{Materials Research and Technology Department, Luxembourg Institute of Science and Technology, 41 rue du Brill, L-4422 Belvaux, Luxembourg}
\newcommand{\BVO}{BiVO$_4$}
\newcommand{\Ag}{A$_\mathrm g$}
\newcommand{\Bg}{B$_\mathrm g$}

\renewcommand{\figurename}{Supporting Fig.}
\renewcommand{\thetable}{S\arabic{table}}
%\preprint{APS/123-QED}

%\title{Spectroscopic signatures of electron-polarons in monoclinic bismuth vanadate}
\title{Supporting information: Raman resonances mediated by excitonic polarons in BiVO$_4$}
% Force line breaks with \\

\author{Georgy Gordeev}
 \affiliation{\Lux{}}
 %\email{georgy.gordeev@uni.lu}
\author{Christina Hill}%
\affiliation{\Lux{}}
\affiliation{\LIST{}}
\author{Angelina Gudima}
\affiliation{\Lux{}}
\author{Stephanie Reich}
\affiliation{\FU{}}
\author{Mael Guennou}
\affiliation{\Lux{}}

\date{\today}% It is always \today, today,
\maketitle
%\newpage
%\section{Line-scan measurement}
\section{Raman scattering cross-section}
\begin{figure}
    \centering
    \includegraphics[width=0.5\linewidth]{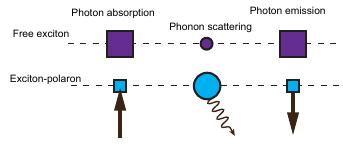}
    \caption{Scheme of resonant Raman scattering processes via exciton and exciton-polaron (EP). Vertical arrows are the photonic transitions and tilted arrows represent scattering by phonons.}
    \label{fig:fey}
\end{figure}

The Raman process mediated by the polaron is expected to have a superior cross-section. A  Feynman diagram is depicted in Figure \ref{fig:fey}, where we illustrate a typical three-step Stokes Raman process. The visible light photon is first absorbed, then phonon is created followed by re-emission of the photon at a lower energy. The size of the vertex represents the interaction strength. For EP the interaction with the photon is weak, but it can be compensated by an enhanced EP-phonon scattering probability, since it is formed by the phonon interactions. The Feynman diagrams can be converted to a perturbation theory equation. We focus on symmetric \Ag{} modes with dominant diagonal tensor elements, $R^{A_\mathrm g}_{jj}$, with $j=a,b,c$. The wavelength dependence of $R^{Ag}_{jj}$ cross-section has contributions from FE and EP \cite{Yu1995}
\begin{equation}
    R^{Ag}_{jj}(E_l) \propto E_l^4 \left| R^{exc}_{jj}(E_l) + R^{EP}_{jj}(E_l) \right|^2
    \label{EQ:Rtot}
\end{equation}
A contribution of a resonant state  $R^{exc,EP}_{jj}$ is frequently described by the third-order perturbation theory, yielding dependence on laser energy $E_l$ as\cite{Yu1995}
\begin{equation}
    R^{Ag}_{i,jj}(E_l) = \frac{M_{photon-i,jj}^2M_{i-ph}}{(E_l - E_{i}^{jj}-i\gamma_i)(E_l - E_{i}^{jj}-E_{ph}-i\gamma_i)}.
    \label{EQ:pert}
\end{equation}
For excitons, where $i = exc,EP$, we find the matrix element $M_{photon-exc,jj}$ in the numerator that represents an interaction between an exciton and a photon, and $j$ indicates polarization of the electric field oriented along $a, b$ or $c$ axis. $M_{exc-ph}$ is the exciton-phonon matrix element. $E_{i}^{jj}$ represents the energy of an exciton with a non-zero transition-dipole moment with light polarized along $j$. This energy matches the $E_{CB}-E_{VB}-E_B^{exc}$, $E_B$ is the binding energy of the exciton, and $E_{ph}$ is the energy of the phonon. The denominator contains so-called incoming and outgoing Raman resonances. \par

%figs1
%\section{Raman resonances of the Ag modes}
\begin{figure}
    \centering
    \includegraphics[width=0.9\linewidth]{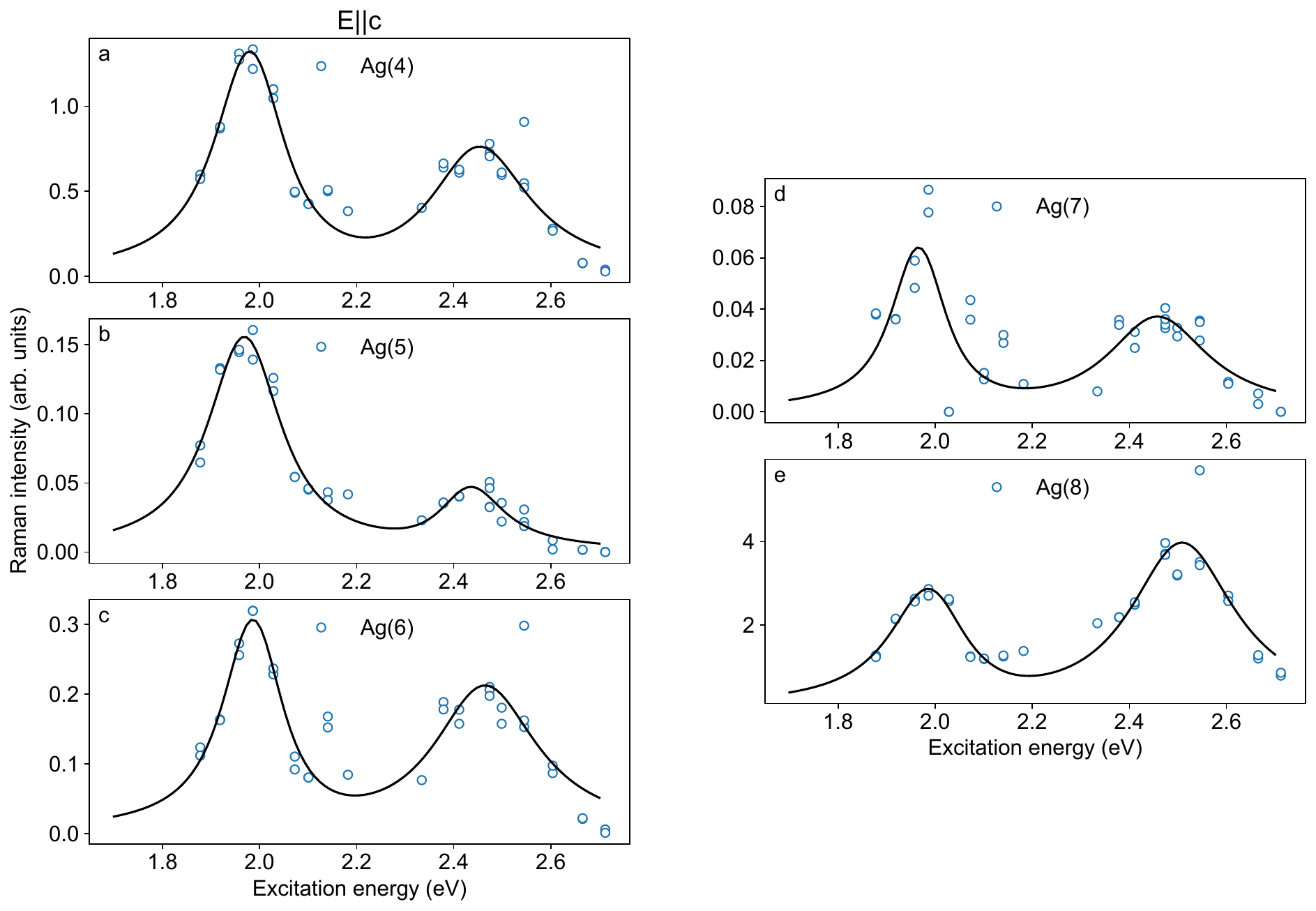}
    \caption{Summary of the resonant Raman profiles from Ag modes excited in the $E\parallel c$ configuration }
    \label{fig:placeholder}
\end{figure}

\section{Optical data}
\begin{figure}
    \centering
    \includegraphics[width=0.9\linewidth]{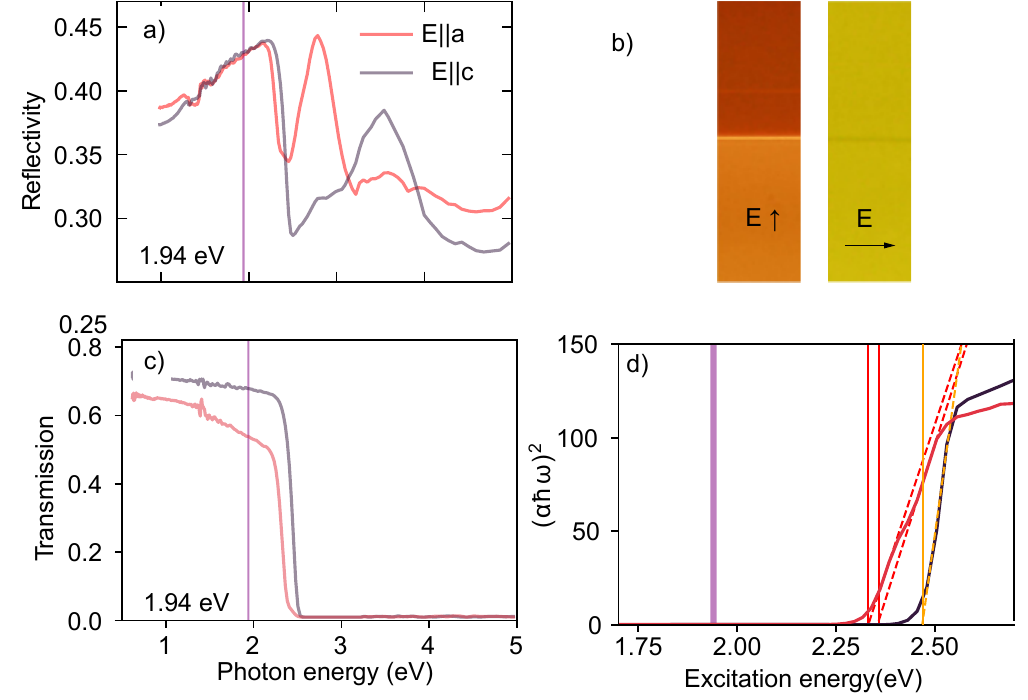}
    \caption{Optical analysis of the anisotropy in \BVO{}. (a) Reflectivity spectra with light polarized along and perpendicular to the c-axis. (b) Optical image of the sample in transmission mode. (c) Transmission spectra. (d) Tauc plots calculated from the transmission and reflectivity data.}
    \label{fig:OPT}
\end{figure}

We now measure the linear optical properties of \BVO{} to disentangle the contributions from the EP and exciton states. The optical spectra were measured in \SI{5}{\nm} steps with light polarized parallel and perpendicular to the $c$ axis using a Lambda 1050+ uv/vis/nir Perkin-Elmer spectrometer, see Figure \ref{fig:OPT}c. Anisotropy in the lowest optical band gap energies visible both in transmission and reflectivity spectra, shown in Figure \ref{fig:OPT}a,b.  We quantify the optical band gaps using Tauc plots shown in Figure \ref{fig:OPT}c, where $(\alpha \hbar \omega)^n$ is plotted versus $\hbar \omega$. The absorption coefficient $\alpha$ is calculated as $\frac{1}{d}log(\frac{1-R}{T})$. For fitting the Tauc plots, we choose $n=2$, the most appropriate for crystalline materials.\cite{Klein2023} The linear fit yields single energy for $E^c_{exc} = \SI{2.47}{\eV}$ and two energies for $E\perp c$, $E^{a,b}_{exc} =  2.33$ and \SI{2.37}{\eV}, that correspond to a and b domains.\cite{Hill2020} These high-energy states manifest themselves in regular way both in optical absorption and resonant Raman, thus we confirm that the high-energy Raman resonance belongs to the excitonic state. \par

\bibliography{main}% Produces the bibliography via BibTeX.